\documentclass[journal=nalefd, manuscript=article]{achemso}

\usepackage[version=3]{mhchem}

\title[\texttt{achemso}] {Enhanced light-matter interaction in graphene-covered
gold nanovoid arrays}

\author{Xiaolong~Zhu}
\affiliation[DTU Fotonik]{Department of Photonics Engineering, Technical University of Denmark, DK-2800 Kongens Lyngby, Denmark}
\alsoaffiliation[CNG]{Center for Nanostructured Graphene (CNG),
Technical University of Denmark, DK-2800 Kongens Lyngby, Denmark}

\author{Lei~Shi}
\affiliation[Fudan University]{Department of Physics, Key Laboratory of Micro- and Nano-Photonic Structures (Ministry of
Education) and State Key Laboratory of Surface Physics, Fudan University, Shanghai 200433, China}

\author{Michael~S.~Schmidt}

\author{Anja~Boisen}

\author{Ole~Hansen}
\affiliation[DTU Nanotech]{Department of Micro and Nanotechnology,
Technical University of Denmark, DK-2800 Kongens Lyngby, Denmark}
\alsoaffiliation[CINF]{Center for Individual Nanoparticle Functionality (CINF), Technical University of Denmark, DK-2800 Kongens Lyngby, Denmark}

\author{Jian~Zi}
\affiliation[Fudan University]{Department of Physics, Key Laboratory of Micro- and Nano-Photonic Structures (Ministry of
Education) and State Key Laboratory of Surface Physics, Fudan University, Shanghai 200433, China}

\author{Sanshui~Xiao}
\affiliation[DTU Fotonik]{Department of Photonics Engineering, Technical University of Denmark, DK-2800 Kongens Lyngby, Denmark}
\alsoaffiliation[CNG]{Center for Nanostructured Graphene (CNG), Technical University of Denmark, DK-2800 Kongens Lyngby, Denmark}
\email{saxi@fotonik.dtu.dk}

\author{N.~Asger~Mortensen}
\affiliation[DTU Fotonik]{Department of Photonics Engineering, Technical University of Denmark, DK-2800 Kongens Lyngby, Denmark}
\alsoaffiliation[CNG]{Center for Nanostructured Graphene (CNG),
Technical University of Denmark, DK-2800 Kongens Lyngby, Denmark}
\email{asger@mailaps.org}

\begin{document}

\begin{abstract}
The combination of graphene with noble-metal nanostructures is currently being explored for strong light-graphene interaction enhanced by plasmons. We introduce a novel hybrid graphene-metal system for studying light-matter interactions with gold-void nanostructures exhibiting resonances in the visible range.
Strong coupling of graphene layers to the plasmon modes of the nanovoid arrays results in significant frequency shifts of the underlying plasmon resonances, enabling more than 30\% absolute light absorption in a single layer of graphene and up to 700-fold enhancement of the Raman response of the graphene. These new perspectives enable us to verify the presence of graphene on gold-void arrays and the enhancement even allows us to accurately quantify the number of layers. Experimental observations are further supported by numerical simulations and perturbation-theory analysis. The graphene gold-void platform is beneficial for sensing of molecules and placing R6G dye molecules on top of the graphene, we observe a strong enhancement of the R6G Raman fingerprints. These results pave the way toward advanced substrates for surface-enhanced Raman scattering (SERS) with potential for unambiguous single-molecule detection on the atomically well-defined layer of graphene.
\end{abstract}

Graphene is an atomic monolayer formed by carbon hexagons, whose extraordinary electrical and optical properties have led to a range of promising optoelectronic devices\cite{Novo,Zhang,Grig}, such as photodetectors\cite{Fang}, optical modulators,\cite{Lium} and ultra-fast lasers\cite{Sunz}. However, all such devices suffer from the inherently weak interaction between pristine graphene and light (2.3\% light absorption at normal incidence), therefore imposing substantial challenges and restrictions for many electro-optical and all-optical applications\cite{Liuj,Enge}. Doped graphene nanostructures which support surface plasmons in the teraherz and infrared regions offer an exciting route to increase the light-graphene interaction by confining the optical fields below the diffraction limit\cite{Manj,Feiz,Chen,Thong,Zhu}. However, graphene is less attractive when the interband loss becomes large, and it effectively mimics a dielectric material in the visible and near-infrared frequencies\cite{Jabl}. One alternative way to enhance the light-graphene interaction in short wavelengths is the combination of graphene with conventional plasmonic nanostructures based on noble metals\cite{Echtermeyer}. These graphene-plasmonic hybrid structures could be beneficial for both fields of investigation: first of all, graphene can influence the optical response of plasmonic structures leading to graphene-based tunable plasmonics\cite{Emani}, and in turn, plasmonic nanostructures can dramatically enhance the local electric field, leading to strong light absorption and Raman signature of graphene layers.

In this Letter, a novel platform based on graphene-covered gold nanovoid arrays (GNVAs) is proposed to enhance the light-matter interaction in graphene-plasmonic hybrid structures. Nanoscale gold-void arrays possess various plasmon modes with significant field enhancement at the cavity voids and rims\cite{Cole}. By tuning the dimensions, the GNVAs exhibit spectrally narrow and wide-angle resonances throughout the visible and near-infrared regimes, which drives applications in surface enhanced Raman scattering (SERS)\cite{Kneipp,Coler,Schmidt}, photovoltaics,\cite{Laln} and sensing\cite{Zhux}. Here we report a dramatic enhancement of the light-matter interaction in the graphene-covered GNVAs with the aid of strongly localized optical fields generated by the GNVAs. The strong interaction in the the graphene-covered GNVAs results in significant frequency shifts of the plasmon resonances, enabling 30\% higher absolute light absorption and up to 700-fold enhancement of graphene Raman signals. The GNVAs are also be applied to the spectroscopy of graphene. In particular, we show that the GNVA allows to easily detect the presence of graphene layers on top of the plasmonic structure and that the resonance shift is sufficient enough to count the actual number of layers. Additionally, the feasibility of graphene-assisted molecule detection by our graphene-plasmonic hybrid structures is discussed. The experimental observations are further confirmed by numerical simulations and theoretical predictions based on perturbation theory. Our results may open a new route to realize strong coupling of electromagnetic fields to graphene layers and to a new generation of compact graphene plasmonic devices.

\begin{figure}[htb]
  \includegraphics[width=5in]{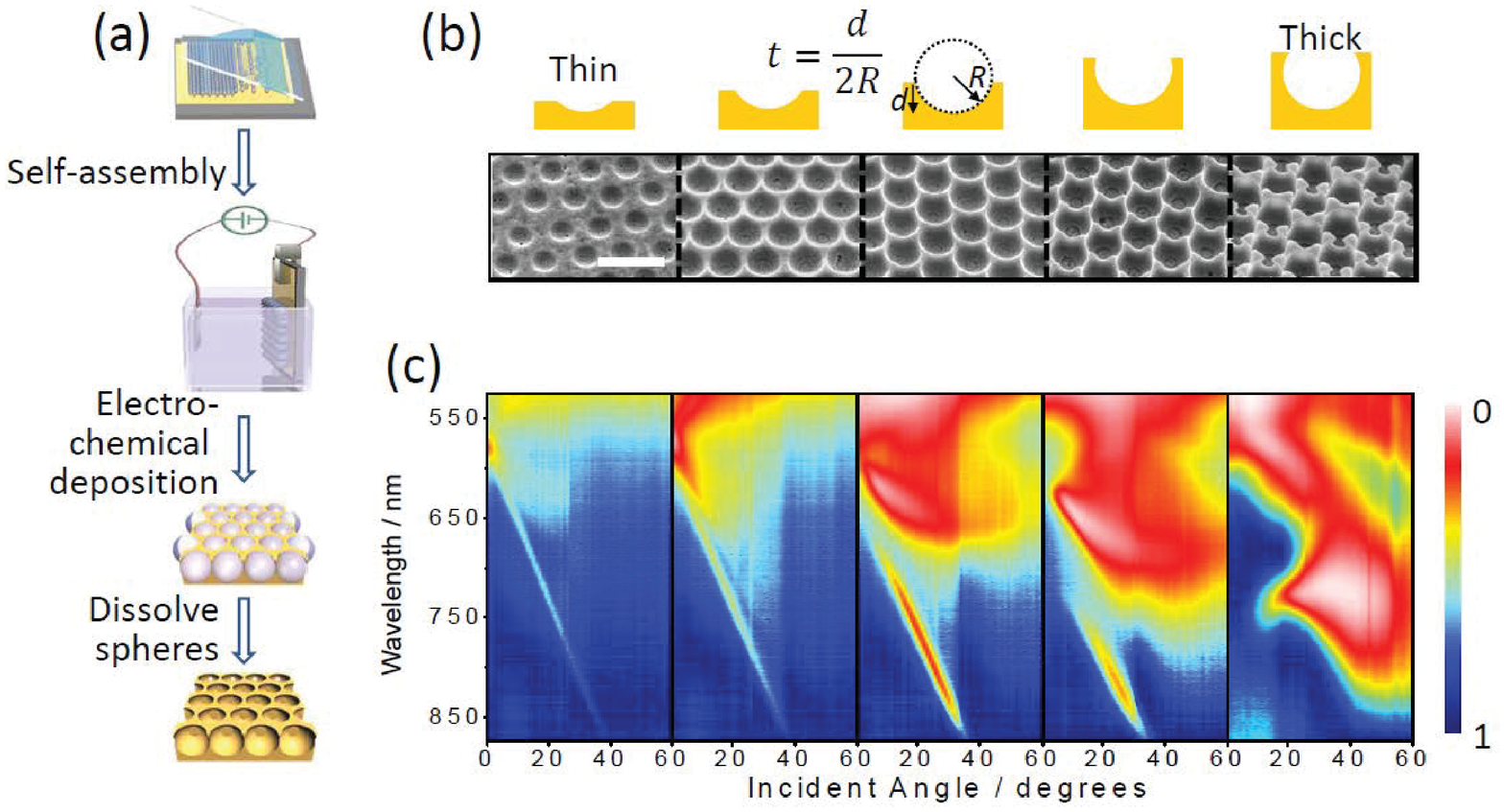}
  \caption{(a) Schematic illustrations of the self-assembly process of monolayer colloidal spheres, the electro-chemical deposition setup, and the nanovoid fabrication procedures. (b) Morphologies (up) and SEM images (down) of 600 nm diameter Au spherical void structures, at various normalized cavity thickness $t=d/2R$, where $d$ is the thickness of the electrochemical deposited gold film and $R$ is the radius of the void. The scale bar is 1 $\mu$m. (c) Measured reflection spectra for the GNVAs with different thicknesses. The corresponding structures are shown in (b).}
  \label{f1}
\end{figure}

Figure \ref{f1}(a) illustrates the fabrication procedure of the GNVAs. Firstly, a monolayer of polystyrene spheres (600 nm in diameter) was prepared on gold-coated substrates using the capillary effect. Electrochemical deposition was then performed on these substrates to obtain templates with morphology-controllable features. After dissolving the polystyrene spheres, structures and shapes ranging from open periodic dish arrays up to fully encapsulated spherical cavities, see Figure \ref{f1}(b), can be realized by carefully controlling the electrochemical deposition time\cite{Zhuxl}. Effectively, the deposition time controls the normalized cavity thickness $t=d/2R$, where $d$ is the thickness of the electrochemical deposited gold film and $R$ is the radius of the void. Figure \ref{f1}(c) shows the measured angle-dependent reflectance for the corresponding structures. Due to the excitation of surface plasmon propagating modes on the surfaces as well as void-plasmon modes localized inside the cavities, strong absorption regions appear in the visible and near-infrared spectra. The enhanced absorption in the optically thin GNVAs relies on delocalized plasmon excitations, which are highly sensitive to the angle of incidence. Conversely, the thick GNVAs exhibiting localized plasmon excitations can be used to absorb light because the localized plasmon resonances can be efficiently excited over a wide range of incidence angles, see Figure \ref{f1}(c). It should be mentioned that the self-assembly method applied here is well suited for fabricating large-area and inexpensive structures, and that by scaling the nanospheres dimensions, the plasmon resonant frequency can be further tuned throughout the visible and infrared regimes.

\begin{figure}[htb]
  \includegraphics[width=5in]{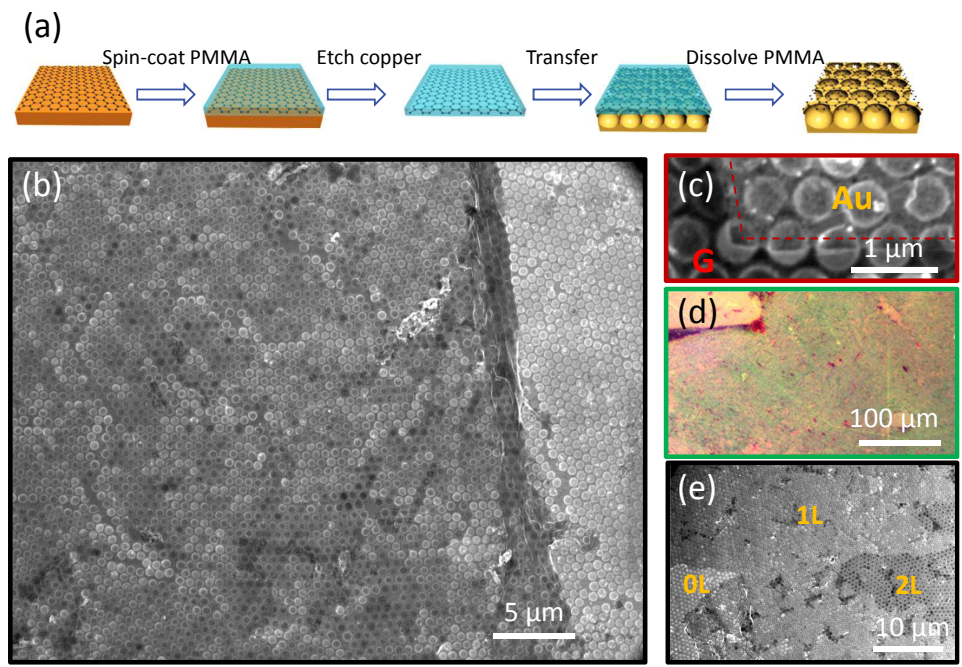}
  \caption{(a) Schematic illustrations of the graphene transfer process. (b) SEM image of a large-area nanovoid array integrated with transferred monolayer graphene. The dark region is covered by graphene. (c) The graphene-covered gold nanovoids where the dashed lines define the edges of the graphene monolayer. (d) Optical microscope image of the nanovoids with monolayer graphene. The part with green color is the region where graphene is located, enabling high-visibility of the graphene sheet. (e) SEM image of a nanovoid array which shows clear boundaries of monolayer and bilayer graphene.}
  \label{f2}
\end{figure}

Monolayer graphene films grown by chemical vapor deposition (CVD) are then wet transferred onto the GNVAs as schematically shown in Figure \ref{f2}(a). In the first step, a poly(methyl methacrylate)(PMMA) film was spin-coated onto the graphene covered copper foil and dried at 170$^{\circ}$C for 30 min. Subsequently, a PMMA/graphene membrane was obtained by etching away the copper foil in a Fe(NO$_{3}$)$_{3}$/H$_{2}$O solution and then transferred to the GNVAs. Finally, we dissolved the PMMA in acetone and cleaned the graphene surface. The scanning electron microscope (SEM) image in Figure \ref{f2}(b) reveals the successful transfer of the CVD monolayer graphene on top of the GNVA. This technique is well suited for large-scale production of the graphene-plasmonic hybrid structures due to the mechanical flexibility of the graphene sheet and the flat interfaces of truncated spherical void arrays.
Traditional plasmonic nanostructures, e.g. split-ring resonators\cite{Sara,Papa}, nanodots\cite{Sche} and nanoholes\cite{Haoq} formed by e-beam lithography or nanopyramids\cite{Wang} fabricated by nanoimprinting were also used for realizing graphene-plasmonic hybrid structures.
For many of these structures, the two-dimensional nature of the graphene film prevents it from fully accommodating to the rapidly varying topography of the structured metallic substrate. The protruding metal structures may create wrinkles and breakages in the on-top transferred graphene layer\cite{Wang}. Therefore, there is a potential for further enhancing the effect of light-graphene interaction by turning to other geometries. Alternatively, the gold nanovoid supports plasmon modes located at its opening, which is ideal for making a seamless and smooth contact with graphene (Figure \ref{f2}(c)). This configuration can provide relatively large interaction area between graphene and the strong electrical field of plasmon resonances. Interestingly, the presence of graphene deposited on the GNVAs becomes immediately visible and its features can be clearly verified, see the green part in Figure \ref{f2}(d). The high color-contrast visualization of graphene layers and potential plasmon-based sensing will be discussed in the following part. Furthermore, the GNVAs with a bilayer graphene (Figure \ref{f2}(e)) are also achieved by repeating the graphene transfer procedure.

\begin{figure}[htb]
  \includegraphics[width=5in]{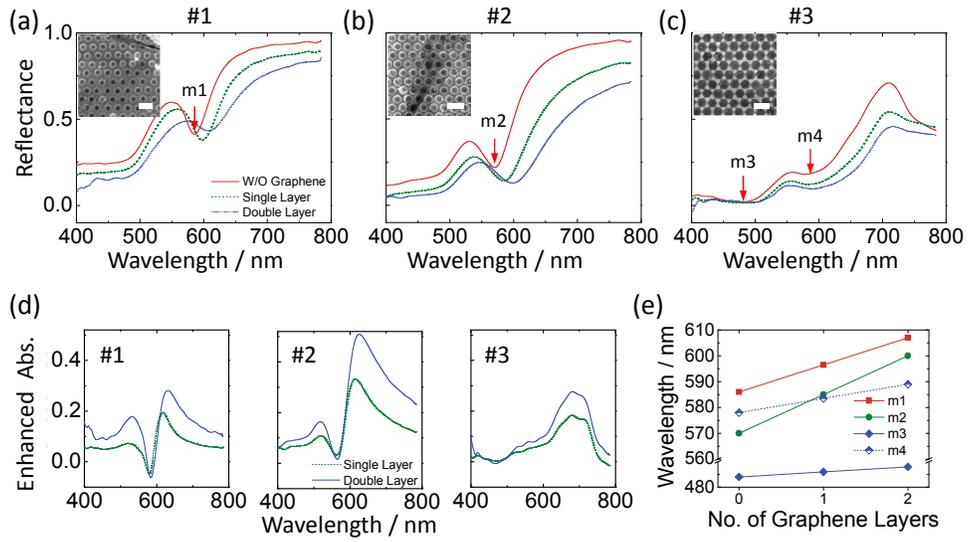}
  \caption{Measured reflection spectra of nanovid arrays before (the solid red lines) and
after deposition of monolayer (the dashed green lines) and bilayer graphene (the dotted blue lines) for the normalized cavity thickness t = 0.1 (a), 0.25 (b) and 0.75 (c), respectively. The red arrows mark the positions of the plasmon-modes for the bare GNVAs without graphene. The insets show the corresponding SEM pictures with scale bars of 1 $\mu$m. (d) Enhanced absolute absorption of the monolayer-(the dotted green lines) and bilayer-graphene (the solid blue lines) covered GNVAs with different normalized cavity thicknesses, which are calculated from (a)-(c). (e) Wavelength shifts for different plasmon modes (mentioned in (a)-(c)) when having different number of graphene layers.}
  \label{f3}
\end{figure}

By controlling the electrochemical deposition time, the size and the shape of the voids can be varied. The insets in Figure \ref{f3}(a)-(c) depict the SEM images of the graphene-plasmonic structures made with different $t$. When structured at subwavelength scales, metals become poor reflectors and they may even turn black\cite{Sondergaard}, which can be largely attributed to absorption triggered by plasmon excitations. Figure \ref{f3}(a)-(c) plot the measured optical reflectance versus wavelength for the GNVAs covered with and without graphene layers when $t=$0.1, 0.25, and 0.75, respectively. We find that the reflection of the bare GNVAs ($R_0$) is suppressed in the regions where surface plasmon resonances are excited. The reflection of a monolayer-graphene-covered GNVAs ($R_1$) with different $t$ show similar trends: a decrease of the reflection is observed while the resonance dips become shallower, broader, and red-shifted. Generally, graphene acts as a lossy dielectric at visible and near-infrared wavelengths. Suppressed reflection and red-shifted plasmonic frequencies are attributed to the graphene losses and presence of graphene. It has been shown that higher refractive index of the surrounding medium results in a longer resonance wavelength\cite{Yux}. The effect is further intensified when having an additional graphene layer: the reflection ($R_2$) and the plasmonic resonance are further decreased and red-shifted, respectively.

The enhanced absolute absorption of the monolayer ($R_0-R_1$) and bilayer ($R_0-R_2$) graphene-covered GNVAs are shown in Figure \ref{f3}(d). The modification of the optical absorption achieved by graphene varies for different GNVAs and can be as high as 30\% for the monolayer and 50\% for the bilayer, confirming that the interaction of light with graphene is strongly enhanced by the plasmonic structures. The average enhanced absorption in the entire visible range of the graphene monolayer (bilayer) covered GNVAs \#1, \#2 and \#3 are 7.1\% (13.7\%), 14.7\% (22.8\%), and 6.4\% (11.0\%), respectively. The phenomenon of enhancing total absorption persists for oblique incidences, as shown in Supplementary Information, Fig. S1. The plasmonic resonance frequencies with respect to the number of graphene layers are shown in Figure \ref{f3}(e). For the monolayer graphene case, a consistent wavelength shift of the m2 plasmon mode up to 15 nm is observed. Particularly, the reflection dips in Figure \ref{f3}(c) experience much smaller wavelength shift than the others, especially for the m3 mode at around 485 nm, whose shift almost cannot be distinguished.

\begin{figure}[htb]
  \includegraphics[width=5in]{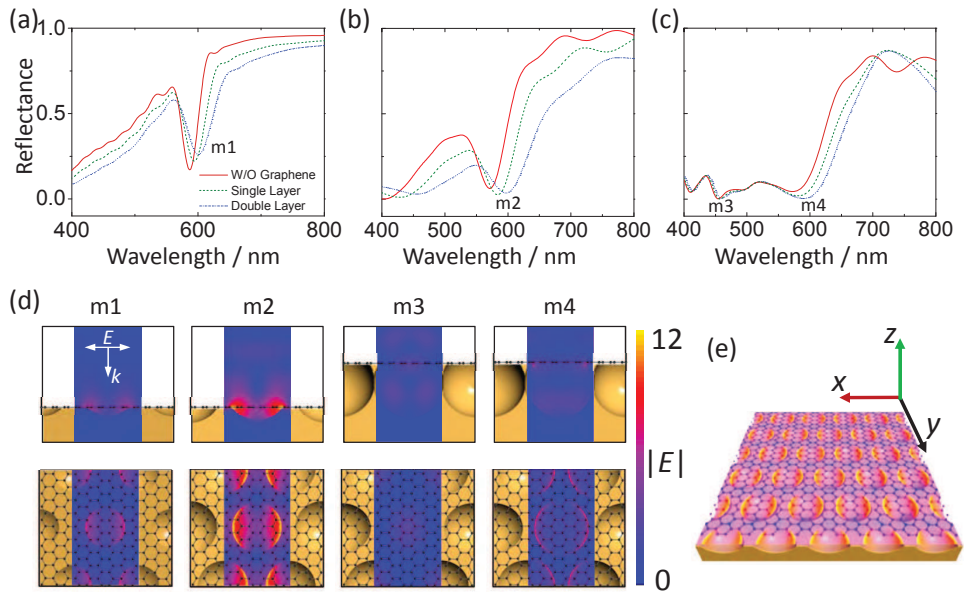}
  \caption{(a)-(c) Simulated reflection spectra of nanovoid arrays before (the solid red lines) and after deposition of monolayer (the dashed green lines) and bilayer graphene (the dotted blue lines). (d) The simulated electrical field distribution of corresponding modes indicated in (a)-(c) for the monolayer-graphene-covered GNVAs. The color bar demonstrates the magnitude of electrical field which is normalized by the incident field. (e) Illustration of spatial coordinates and the nanovoid array \#2 covered by a monolayer graphene. The plasmonic resonance exhibits a strong optical-field overlap with the added graphene layer.}
  \label{f4}
\end{figure}

To explain these experimental observations, full-wave numerical electromagnetic simulations based
on a finite-integration technique (CST MICROWAVE STUDIO) are performed. The refraction index of graphene layers ($n_{g}$) stacked in the natural graphite order in the visible range is governed by $n_{g}=3.0+C\frac{\lambda}{3}i$\cite{Brun}. The effective thickness of CVD-graphene is chosen as $t_{g}$ = 1 nm (see Supplementary Information, Fig. S2). The simulated results in Figure \ref{f4}(a-c) show remarkable agreement with the experimental results. Figure \ref{f4}(d) illustrates the normalized electrical field distribution for the corresponding modes excited in GNVAs covered by the monolayer graphene. The field distribution in the $xz$-plane is shown in the upper four diagrams of Figure \ref{f4}(d) while the lower ones show the corresponding distribution for the cross section in the $xy$-plane propagating in the middle of the graphene layer. As shown in Figure \ref{f4}(d), the m2 plasmon mode of the structure \#2 possesses the strongest optical field and the largest optical overlap with the graphene layer (as illustrated in Figure \ref{f4}(e)), indicating the strongest light-graphene interaction and the most considerable resonance shift as well as enhanced absorption near the resonant frequency.
Actually, when the thickness $t$ of the spherical nanovoid cavity reduces gradually, the localized mode captured in the voids turns into the exposed propagating (delocalized) surface plasmon mode and the field becomes closer to the upper surface where graphene is located too. In contrast, the m1 mode of the structure \#1 possesses a relatively large optical overlap but small value of the field, and the m4 mode of the structure \#3 only has a strong optical field localized at the rim of the nanovoid openings, whose resonance shift and enhanced absorption near the resonant frequency are moderate. Specifically, the m3 plasmon mode of the structure \#3 experiences a negligible wavelength shift and enhanced absorption because the optical field for this mode is strongly localized and captured inside the void and the optical overlap with graphene is minimal. The above phenomenon keeps unchanged for the oblique incidence as shown in Supplementary Information, Fig. S1, that the structure \#3 shows smaller absorption enhancement than that of the structure \#2.

Since the addition of a graphene layer only causes a weak perturbation of the optical response, we may conveniently apply perturbation theory to analyze the frequency shift $\Delta\omega$ due to a small change $\Delta\epsilon$ in the dielectric function. Accounting for the dispersive properties of the plasmonic substrates, we have
\begin{equation}
\frac{\Delta\omega}{\omega}=-\frac{\int \Delta\epsilon \big|E\big|^2\, d^{3}x}{2\int \big(\epsilon(\omega)+\tfrac{1}{2}\omega\frac{\partial\epsilon(\omega)}{\partial\omega}\big)\big|E\big|^2\,d^{3}x}+O(\Delta\epsilon^{2})
\end{equation}
where $E$, $\omega$, $\epsilon(\omega)$ are the electric field, the eigenfrequency, and the dielectric function associated with the unperturbed structure\cite{Rama}.
In the language of quantum mechanics, the result states that the shift is just proportional to the expectation value of the perturbing potential. For the shift to be significant, the extend of the region in which the perturbation occurs is important and likewise, the field intensity has to be large too.
Furthermore, on the scale of the atomic-thin graphene the optical field appears close to uniform and the spatial dependency inside the integrals in Eq.~(1) can be safely ignored. In the case of multi-layer graphene, the resonance shift in this way becomes directly proportional to the number of layers with the constant of proportionality given by the resonance shift associated with a single layer, see Figure \ref{f3}(e). The shift can easily be 10\,nm per layer, depending on the underlying plasmonic nanostructure. For a higher number of layers we anticipate an interplay between the total thickness and the characteristic decay length of the localized plasmon field. The wavelength shift decreases exponentially when increase the number of graphene layers and the total shift saturates when approaching 10 layers of graphene (see Supplementary Information, Fig. S3).

\begin{figure}[htb]
  \includegraphics[width=5in]{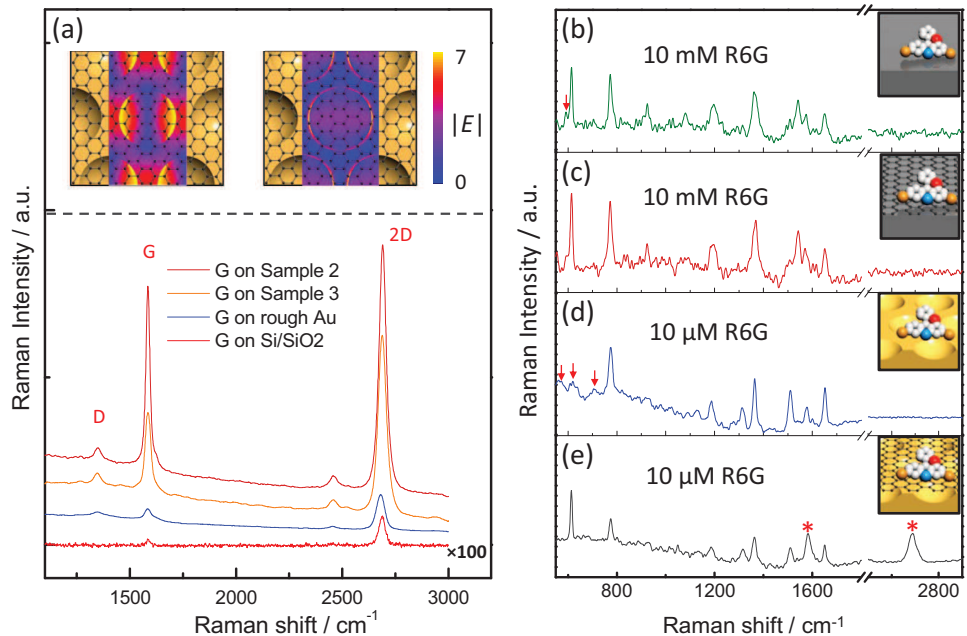}
  \caption{(a) Evolution of the Raman spectra of monolayer graphene on different substrates. The spectra
are shifted vertically for clarity. The insets are the simulated electrical field distributions at 532 nm for the GNVAs \#2 (left) and \#3 (right) covered by monolayer graphene. (b)-(e) Raman scattering spectra of R6G on different substrates. Red arrows here point to additional and irreproducible peaks in the normal Raman spectra of R6G. "$\ast$" in (e) mark the G-band and 2D-band of monolayer graphene. All Raman spectra are measured under the same conditions and are shown on the same intensity scale, except for the concentration of R6G, which is labeled in corresponding spectra. It should be noted that the characteristic Raman shift peaks of graphene are too weak to be detected in (c).}
  \label{f5}
\end{figure}

As a result of the enhanced light-graphene interaction, the Raman scattering of graphene can be modified tremendously via the GNVAs.
Figure \ref{f5}(a) depicts typical SERS spectra of the single-layer graphene using 532 nm laser excitation for the structures \#2 and \#3. The Raman spectra of the monolayer graphene on a rough gold film (a traditional SERS substrate, see Supplementary Information, Fig. S4) and a bare SiO$_{2}$/Si substrate are also measured for comparison. Prominent features with Raman shifts of the G peak around 1580 cm$^{-1}$ and 2D peak around 2690 cm$^{-1}$ are observed for all samples. The 2D peak exhibits a single Lorentzian shape, which is the signature of a monolayer graphene with a single-band electronic dispersion. The low-intensity peak around 1350 cm$^{-1}$ is attributed to the D band of carbon, suggesting a low presence of defects in the used CVD graphene as well as few adsorbents remaining from the graphene transfer process.
Although the maximum enhancement can only be achieved when the plasmonic frequency is well matched to the excitation (Figure \ref{f5}(a)), the enhancement of the Raman scattering of graphene on GNVAs is still considerable when compared with that of the graphene on the SiO$_{2}$/Si substrate. For instance, the Raman scattering of both the G
and 2D bands is enhanced dramatically when graphene is optical pumped on the GNVAs and the enhancement factor can reach 700 folds.

In SERS, the electromagnetic enhancement originates from the excitation of strongly localized electrical fields with incident light. The plasmonic enhancement factor is commonly estimated as\cite{Garc,Kneipp}
\begin{equation}
\gamma\approx\frac{|E_{t}|^{4}}{|E_{i}|^{4}}
\end{equation}
where $E_{i}$ is the incident field, and $E_{t}$ is the total field at the molecule location.
The insets of Figure \ref{f5} show the corresponding simulated electrical field distribution at 532 nm in the monolayer-graphene-covered GNVAs. Because of the special geometry configurations and the symmetry properties of the truncated spherical arrays, graphene is located at the plane with the strongest field intensity and have an optimized overlap with it (Figure \ref{f4}(d)), leading to the huge enhancements of the Raman scattering. The largest enhancement of electrical field intensity in the structure \#2 reaches 7, giving an approximate 2400 electromagnetic enhancement for SERS. This correlates to the 700 folds SERS enhancement in the experiment.

Although a rough surface is widely accepted to be a critical premise where a strong SERS happens, for many analytical situations a flat SERS surface is highly preferable. The unique property of graphene, i.e., a perfect 2D crystal which is atomically smooth, atomically thin and chemically inert, have already made it to be an ideal building block of a flat SERS surface\cite{Xuw,XuwPNAS}. The combined structure consisting of the GNVA and the flat on-top graphene is a natural candidate for improving SERS detection. In order to seek possible superiorities of these graphene-covered GNVAs for SERS, a uniform layer of rhodamine 6G (R6G) was decorated on GNAVs (structure \#2) with and without graphene. Typical Raman spectra of R6G in the GNVA with and without graphene are shown in Figure \ref{f5}(d-e). As a comparison, Raman signals on SiO$_{2}$/Si and graphene-SiO$_{2}$/Si substrates are also measured and shown in  Figure \ref{f5}(b-c). One can see that the SERS intensities in Figure \ref{f5} are almost in the same order, but the substrates with GNVAs in Figure \ref{f5}(d-e) give an approximate SERS enhancement of 10$^{3}$ when the concentration of the R6G is taken into account.
For the SiO$_{2}$/Si and graphene-SiO$_{2}$/Si substrates, Raman signals provide unclean baselines that suffer from photo-induced damages such as photo-carbonization and photo-bleaching when the concentration is relatively high.
The graphene-covered GNVA shows highly consistent results with a clear baseline in Figure \ref{f5}(e), while the GNVA without graphene possesses some inconsistent peaks (with shifted or some new emerging features, which are irreproducible and may be attributed to photochemical effects), as seen in Figure \ref{f5}(d). According to the above-mentioned key points, the unique graphene-covered GNVAs where graphene acts as an atomic-thin and seamless adhesion layer with chemical inertness could be an ideal substrate for SERS applications. Moreover, the graphene layer can be suspended between the void rims with a small proportion directly attaching the metal, and the molecules can be completely isolated from the metallic substrate. Therefore, the photo-induced damages like photo-carbonization and photo-bleaching as well as the chemical absorption and deformation may be highly prevented.

In conclusion, we have demonstrated that the interaction of light with graphene can be greatly enhanced when graphene is in a strong proximity to metallic nanostructures. When placing graphene on top of gold-void arrays, we observe a significant shift of the underlying plasmonic resonances. Furthermore, for a single layer of graphene the light absorption can reach 30\% as opposed to the 2.3\% in pristine graphene itself. This pronounced change in absorption is clearly visible and it enables easy detection of the graphene sheet and an unambiguous quantification of the number of layers. With the aid of the gold-void arrays, we find 700-fold enhancement of the characteristic Raman signal of graphene. Furthermore, the strong Raman enhancement carries over to the situation where additional R6G dye molecules are placed on top of the graphene, thus demonstrating a potential for future SERS platforms for unambiguous single-molecule detection on the atomically well-defined layer of graphene.


\begin{acknowledgement}
This work was partly supported by the Catalysis for Sustainable Energy Initiative Center (CASE), funded by the Danish
Ministry of Science, Technology and Innovation. The Center for Nanostructured Graphene (CNG) is sponsored by the Danish National Research Foundation, Project DNRF58. The Center for Individual Nanoparticle Functionality CINF is funded by the Danish National Research Foundation, Project DNRF54.
\end{acknowledgement}

%


\providecommand*\mcitethebibliography{\thebibliography}
\csname @ifundefined\endcsname{endmcitethebibliography}
  {\let\endmcitethebibliography\endthebibliography}{}

\subsection{Figures}

\end{document}